\documentclass[twocolumn,twoside,slac_two]{revtex4}
\usepackage{graphicx}
\usepackage{fancyhdr}
\usepackage{url}
\usepackage[justification=raggedright, singlelinecheck=false, compatibility=false]{caption}  
\usepackage{subcaption}
\usepackage{amssymb}

 \newcommand{\hi}{$\mathrm{H\,\scriptstyle{I}}$}

\begin{document}

\title{A Method for Exploring Systematics Due to Galactic Interstellar Emission Modeling: Application to the \textit{Fermi} LAT SNR Catalog}

%

\author{F. de Palma}
\affiliation{ INFN Sezione di Bari, 70126 Italia}
\author{T. J. Brandt} 
\affiliation{NASA Goddard Space Flight Center, Greenbelt, MD 20771, USA}
\author{G. Johannesson}
\affiliation{Science Institute, University of Iceland, Dunhaga 3, 107 Reykjavik, Iceland}
\author{L. Tibaldo}
\affiliation{Kavli Institute for Particle Astrophysics \& Cosmology, SLAC National Accelerator Laboratoty, USA}
\author{on behalf of the \textit{Fermi} LAT collaboration.}



%

\begin{abstract}
Galactic interstellar emission contributes substantially to \textit{Fermi} LAT
observations in the Galactic plane, where the majority of Supernova Remnants
(SNRs) are located. We have developed a method to explore some systematic
effects on SNRs' properties caused by interstellar emission modeling. We created
eight alternative Galactic interstellar models by varying a few input parameters
to GALPROP, namely the height of the cosmic ray
propagation halo, cosmic ray source distribution in the Galaxy, and atomic
hydrogen spin temperature. We have analyzed eight representative candidate SNRs
chosen to encompass a range of Galactic locations, extensions, and spectral
properties using the eight different interstellar emission models. The models
were fitted to the LAT data with free independent normalization coefficients for
the various components of the model along the line of sight in each region of
interest. We will discuss the results and compare them with those obtained with
the official LAT 
interstellar emission model.
\end{abstract}

\maketitle

\thispagestyle{fancy}


\section{INTRODUCTION}

Galactic interstellar $\gamma$-ray emission is produced through interactions of high-energy cosmic ray (CR) hadrons and leptons with interstellar gas via nucleon-nucleon inelastic collisions and electron Bremsstrahlung, and with low-energy radiation fields, via inverse Compton (IC) scattering. Such interstellar emission accounts for more than $60\%$ of the photons detected by the \textit{Fermi} Large Area Telescope (LAT) and is particularly bright toward the Galactic disk.

In this paper, we present our ongoing effort to explore the systematic uncertainties due to the modeling of Galactic interstellar emission in the analysis of \textit{Fermi} LAT sources, with particular emphasis on its application to the $1^{st}$ \textit{Fermi} LAT Supernova Remnant (SNR) Catalog. We compare the results of analyzing sources with eight alternative interstellar emission models (IEMs), described in Section~\ref{iems}, to the source parameters obtained with the standard model in Section~\ref{SNRtests}. In Section~\ref{appl} we discuss the future application of this method to the SNR Catalog.


\section{INTERSTELLAR EMISSION MODELS}\label{iems}
In order to estimate the systematic uncertainty inherent in the choice of standard interstellar emission model (IEM) in analyzing a source, we have developed eight alternative IEMs. By comparing the results of the source analysis using these eight alternative models to the standard model, we can approximate the systematic uncertainty therefrom.

\subsection{The Standard IEM}
The standard IEM for \textit{Fermi} LAT data analysis was developed by the collaboration using the simple assumption that energetic CRs uniformly penetrate all gas phases in the interstellar medium. Under this assumption, the Galactic interstellar $\gamma$-ray intensities can be modeled as a linear combination of gas column densities and an inverse Compton (IC) intensity map as a function of energy. The gas column densities are determined from emission lines of atomic hydrogen~(\hi{})\footnote{\hi{} column densities are extracted from the radio data using a uniform value for the spin temperature ($200$\,K).} and CO, the latter a surrogate tracer of molecular hydrogen, and from dust optical depth maps used to account for gas not traced by the lines. To account for a possible large scale gradient of CR densities, the gas column density maps were split into 6 Galactocentric rings using the emission lines' Doppler shift. The IC map is obtained using GALPROP\footnote{The GALPROP code has been developed over several years, starting with, e.g. \cite{1998ApJ...493..694M} and \cite{1998ApJ...509..212S}.} 
to reproduce the direct CR measurements with a realistic 
model of the Galactic interstellar radiation field (ISRF), as was done in \cite{PorterICGalRidge}.
To account for some extended remaining residuals, notably Loop~I  \cite{CasandjianLoopI} and the so-called \textit{Fermi} bubbles \cite{SuFinkbeiner2010}, the standard IEM includes them as additional templates. 


This IEM, along with sources in the 2FGL Catalog \cite{2fgl_paper} and an isotropic intensity accounting for the extragalactic $\gamma$-ray and instrumental backgrounds, was fit to two years of LAT data. This yielded best fit values of the linear combination coefficients which can be interpreted as gas emissivities in the various Galactocentric rings and a renormalization for the IC model as a function of energy, as well as the spectrum of the isotropic component.  The ratio of CO to twice the \hi{} emissivities and dark gas to \hi{} emissivities are proportional to the CO-to-H$_2$ and dust-to-gas ratios, respectively, under our simple assumption. 
Another explanation of this method of decomposing the $\gamma$-ray sky to create the standard model may be found in, e.g. \cite{2011ApJ...741...44K}.
The standard IEM is distributed as a cube summed over the components which predicts the intensities of Galactic interstellar $\gamma$-ray emission in a grid of directions and energies with its accompanying isotropic model. Further description and details are available at the \textit{Fermi} Science Support Center\footnote{http://fermi.gsfc.nasa.gov/ssc/data/access/lat/\newline Model\_details/Pass7\_galactic.html}.



\subsection{Alternative IEMs}
To explore the uncertainties related to the modeling of interstellar emission we generated eight alternative IEMs, in particular probing key sources of systematic uncertainties by:
\begin{itemize}
 \item adopting a different model building strategy from the standard IEM, resulting in different gas emissivities, or equivalently CO-to-H$_2$ and dust-to-gas ratios, and including a different approach for dealing with the remaining extended residuals;
\item varying a few important input parameters for building the alternative IEMs: atomic hydrogen spin temperature ($150$\,K and optically thin), CR source distribution (SNRs and pulsars), and CR propagation halo heights (4 kpc and 10 kpc);
\item and allowing more freedom in the fit by separately scaling the inverse Compton emission and \hi{} and CO emission in 4 Galactocentric rings.
\end{itemize}

The work in \cite{2012ApJ...750....3A}, using the GALPROP CR propagation and interaction code,  was used 
as a starting point for our model building strategy.
The GALPROP output intensity maps associated with \hi{}, CO, and IC are then fit simultaneously with an isotropic component and 2FGL sources to 2~years of \textit{Fermi} LAT data in order to minimize bias in the a priori assumptions on the CR injection spectra and the proton CR source distribution. The intensity maps associated with gas were binned into four Galactocentric annuli ($0-4$\,kpc, $4-8$\,kpc, $8-10$\,kpc and $10-30$\,kpc). 
The spectra of all intensity maps were individually fit with log parabolas to the data, to allow for possible CR spectral variations between the annuli for all  \hi{} and CO maps while the IC fit accounts for spectral variations in the electron distribution. 
We also included in the fit an isotropic template and templates for Loop I \cite{CasandjianLoopI} and the \textit{Fermi} bubbles \cite{SuFinkbeiner2010}. The template for Loop I is based on the geometrical model of \cite{Wolleben2007} while bubbles are assumed to be uniform with edges defined in spherical coordinates by $R = R_0 |\sin\theta|$, where $\theta$ is the polar angle.

\begin{figure}[h!]
\begin{center}      
\includegraphics[clip=true, width=1\columnwidth,  trim=0.1in .1in 0.1in .0in]{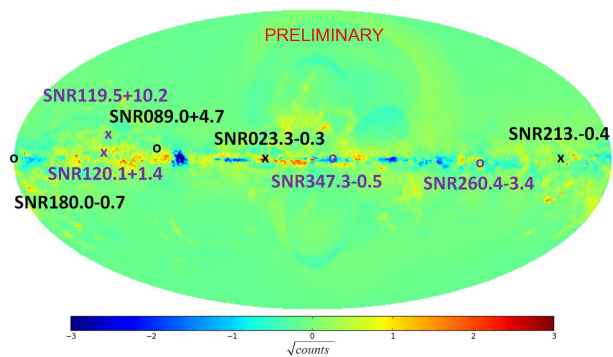}
\end{center} 
\caption{\label{fig:snr_loc} 
The position of the eight candidate SNRs used in this analysis are overlaid on a map of relative difference between the standard IEM and one of the alternative models. The alternative model selected for this image has a Lorimer source distribution, a halo height of $4$\,kpc and a spin temperature of $150$\,K. We plot the difference between the models' predicted counts divided by the square of the sum of the predicted counts so the map is in units of sigma. The hardness of the SNRs' spectra is in two categories: hard (purple) and soft (black). SNRs detected as extended with \textit{Fermi} are shown as circles while point-like are shown as crosses.}
\end{figure}

\cite{2012ApJ...750....3A} explored some systematic uncertainties by varying input parameters. 
The \hi{} spin temperature, CR source distribution, and CR propagation halo height were found to be among those parameters which have the largest impact on the $\gamma$-ray intensity. The values adopted in this study to generate the eight alternative IEMs were chosen to be reasonably extreme; we note that they do not reflect the full uncertainty in the input parameters. 
Separately scaling the \hi{} and CO emission in rings and the IC emission permits the alternative IEMs to better adapt to local structure when analyzing particular source regions.
Figure~\ref{fig:snr_loc} shows the relative difference between the standard model and one of the alternative models (Lorimer CR source distribution with a $4$\,kpc halo height, and $150$\,K \hi{} spin temperature). Differences are particularly large along the Galactic plane, where SNRs are located. 

Finally, we note that this strategy for estimating systematic uncertainty from interstellar emission modeling does not represent the complete range of systematics involved. In particular, we have tested only one alternative method for building the IEM, and the input parameters do not encompass their full uncertainties. Further, as the alternative method differs from that used to create the standard IEM, the resulting uncertainties will not bracket the results using the standard model. 
The estimated uncertainty does not contain other possibly important sources of systematic error, including uncertainties in the ISRF model, simplifications to Galaxy's geometry, small scale non-uniformities in the CO-to-H$_2$ and dust-to-gas ratios and \hi{} spin temperature non-uniformities, and underlying uncertainties in the input gas and dust maps. 
While the resulting uncertainty should be considered a limited estimate of the systematic uncertainty due to interstellar emission modeling, rather than a full determination, it is critical for interpreting the data, and this work represents our most complete and systematic effort to date.

\section{ESTIMATING IEM SYSTEMATICS}\label{SNRtests}

\subsection{Analysis Method}

We developed this method for estimating the systematics from the interstellar emission model using eight candidate SNRs chosen to represent the range of spectral and spatial SNR characteristics in high and low IEM intensity regions. Figure~\ref{fig:snr_loc} shows the candidate SNRs' location on the sky, illustrating their range of Galactic longitude. The color indicates those candidates with a hard or soft index and the shape the extension (pointlike or extended). The SNR candidates are overlaid on a map of the relative difference between the standard IEM and one of the alternative IEMs described in Section~\ref{iems}.

We use the same analysis strategy to obtain all SNR candidates' \textit{Fermi} LAT parameter values with both the standard and all eight alternative IEMs on $3$\,years of P7\_V6SOURCE data \cite{pass7_paper} in the energy range $1-100$\,GeV. 
We applied the standard binned likelihood method\footnote{The standard \textit{Fermi} LAT analysis description and tools can be found here: http://fermi.gsfc.nasa.gov/ssc/data/analysis/ .}, treating sources as follows. 
For each of the eight candidate SNRs, an extended source initially of the radio size and with a power law (PL) spectral model either replaces the closest non-pulsar 2FGL source \cite{2fgl_paper} within the radio size or is positioned as a new source at the location determined from radio observations \cite{green_cat}. All other 2FGL sources within the radio size which are not pulsars are removed from the source model. We fit the centroid and extension of the SNR candidate disk as well as the normalization and PL index for the source of interest and the five closest background sources within $5^{\circ}$ with a significance of $\gtrsim 4\,\sigma$. This procedure balances the number of degrees of freedom with convergence and computation time requirements.

\begin{figure}[h!]
\centering
\includegraphics[clip=true, width=.75\columnwidth, trim=0.1in .3in 0.1in .4in]{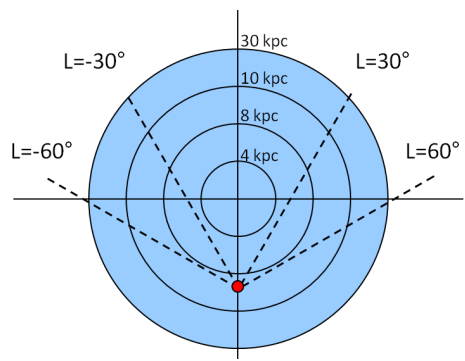} 
\caption{\label{fig:rings} Schematic representation of the \hi{} and CO rings used for the split alternative IEMs crossed by lines of sight at various Galactic longitudes. The ring boundaries lie at $0, 4, 8, 10, 30$\,kpc from the Galactic center while the red dot marks the sun position at $8.5$\,kpc. The figure is not to scale.}
\end{figure} 

\begin{figure*}
        \centering
        \begin{subfigure}[b]{1\columnwidth}
                \centering
                \includegraphics[width=1\columnwidth]{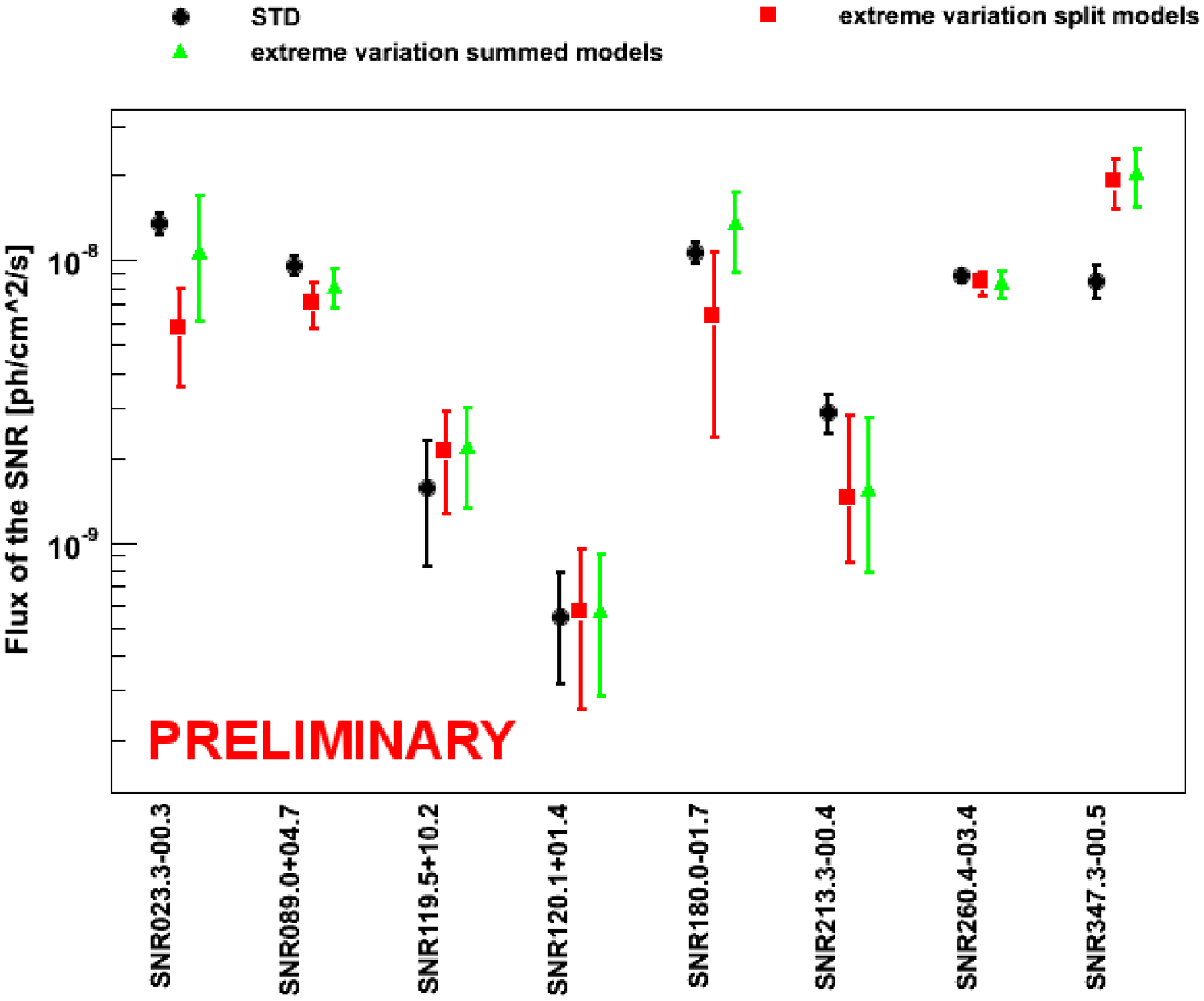}
               \caption{Flux for the eight candidate SNRs' from $1-100$\,GeV.}
                \label{fig:snr_comparisons_flux}
        \end{subfigure}%
        ~ 
        \begin{subfigure}[b]{1\columnwidth}
                \centering
                \includegraphics[width=1\columnwidth]{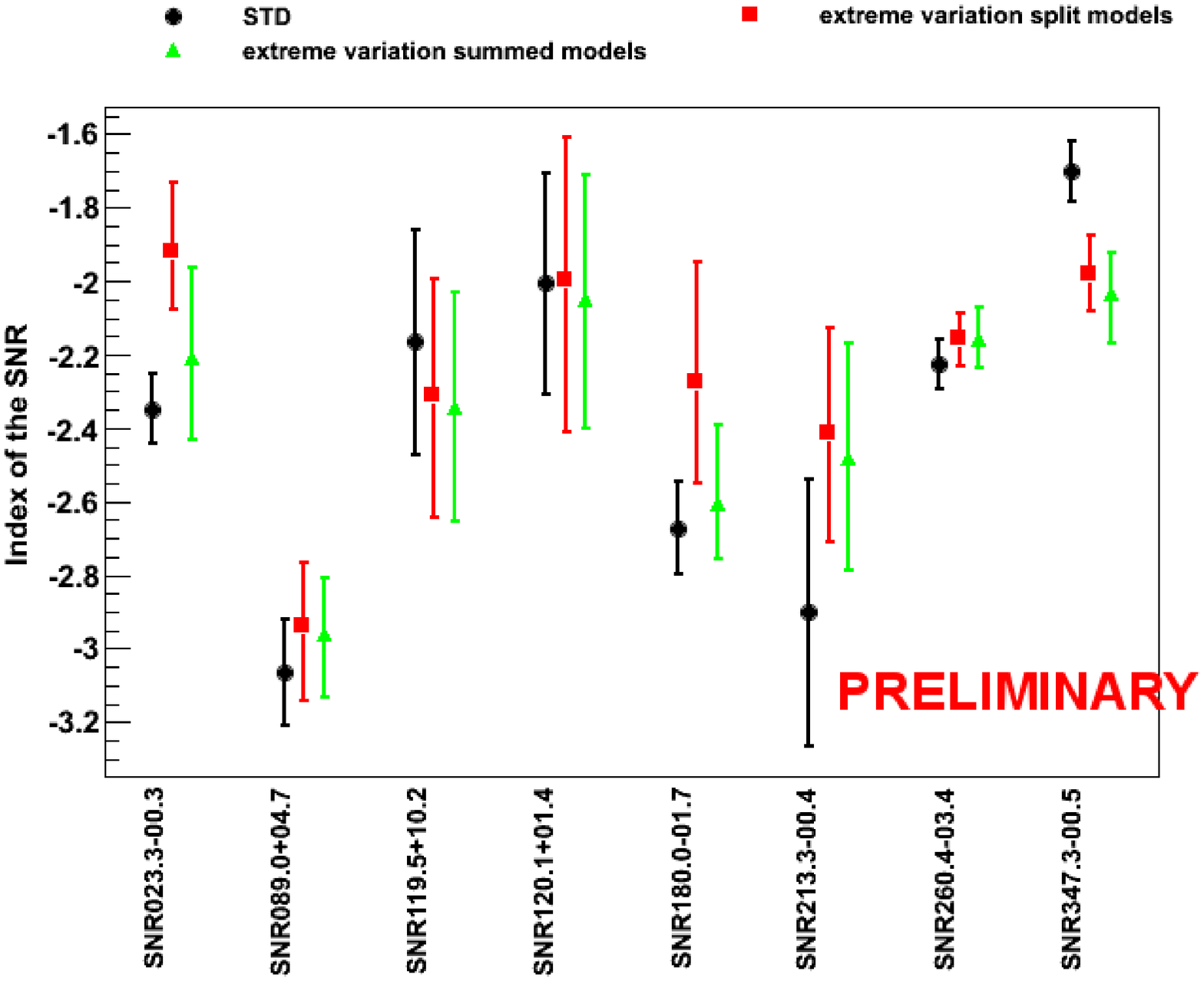}
                \caption{Index for the eight SNR candidates from $1-100$\,GeV.}
                \label{fig:snr_comparisons_index}
        \end{subfigure}
        \caption{Results for each candidate SNR, averaging over the eight alternative IEMs separately for split (red) and summed (green) component models compared to the standard model solution (black). The error bars for results using the alternative IEMs show the maximal range of the values given by the $1\,\sigma$ statistical errors.}\label{fig:snr_comparisons}
\end{figure*}

To generate results for the source of interest with each diffuse model, we fit the sources' model to the data with either the standard model or one of the eight alternative IEMs. In the case of the standard \textit{Fermi} LAT IEM, we allow the normalization to vary and fix the accompanying standard isotropic model's normalization. For each of the eight alternative models, we use the corresponding isotropic model fixed to its value resulting from the fit to the all-sky data (see Section~\ref{iems}). To better understand the effect of allowing freedom in the \hi{} and CO rings, we fit the alternative models in two ways: either with the rings' normalizations free (''split'' models) or with the rings summed together, as given by the all-sky fit (see Section~\ref{iems}), and only the total normalization free (''summed'' models). The summed alternative IEMs are thus closer to the standard IEM. For the split alternative IEMs, as shown in Figure~\ref{fig:rings}, not all rings are crossed by all lines of sight. We thus 
fit only the two innermost  \hi{} and CO rings crossed by the line of sight to our region of interest. The IC template is also free to vary while the isotropic component remains fixed. 
\subsection{Results for SNRs' IEM Systematics}

To compare the results obtained using the eight alternative IEMs with the standard model results, we average each parameter's eight values from the alternative IEMs. Figure~\ref{fig:snr_comparisons} shows the values for the flux and index from fitting the data with the alternative IEMs with the rings either split or summed. These are then plotted along with the standard model results for all eight SNR candidates studied. We conservatively represent the allowed parameter range with error bars showing the maximal range for the alternative IEMs $1\,\sigma$ statistical errors. 

Figure~\ref{fig:snr_comparisons} shows that the variation in value of the best fit parameters obtained with the alternative IEMs is larger than the $1\,\sigma$ statistical uncertainty. The impact of changing the IEM on the source's parameters depends strongly on the source's properties and location. As expected, the parameter values for the source of interest are generally closer to the standard model results for the alternative IEMs with components summed rather than split. In many cases, the allowed parameter range represented by the $1\,\sigma$ statistical errors for each of the alternative IEMs is larger with the components split than summed. 
Also as noted earlier, the alternative IEM results do not as a rule bracket the standard model solution. 
We observe that some of the largest differences between the standard and alternative IEM results for a single source are frequently associated with sources coincident with templates accounting for remaining residual emission in the standard IEM (Section~\ref{iems}). 

SNR G347.3-0.5 proves an interesting source for understanding the impact nearby source(s) can have on this type of analysis. In particular, our automated analysis finds a softer index and a much larger flux for SNR G347.3-0.5 than that obtained in a dedicated analysis \cite{snr347_paper}. Since the best fit radius ($0.8^{\circ}$) is larger than that the X-ray data indicates
($0.55^{\circ}$), the automated analysis's disk encompasses nearby sources that are only used in the \cite{snr347_paper} model. Including this additional emission also affects the spectrum, making it softer in this case than that found in the dedicated analysis. 
Given \textit{Fermi} LAT's both increasing point spread function and number of sources with decreasing energy as well as the predominance of diffuse emission at lower energies, 
we note that nearby sources may play a greater role if extending this method below the $1$\,GeV minimum energy examined here. 

\subsection{IEM Input Parameter Comparison}

\begin{figure}
\centering
\includegraphics[clip=true, width=1\columnwidth, trim=0.5cm 0cm 0cm 0.5cm]{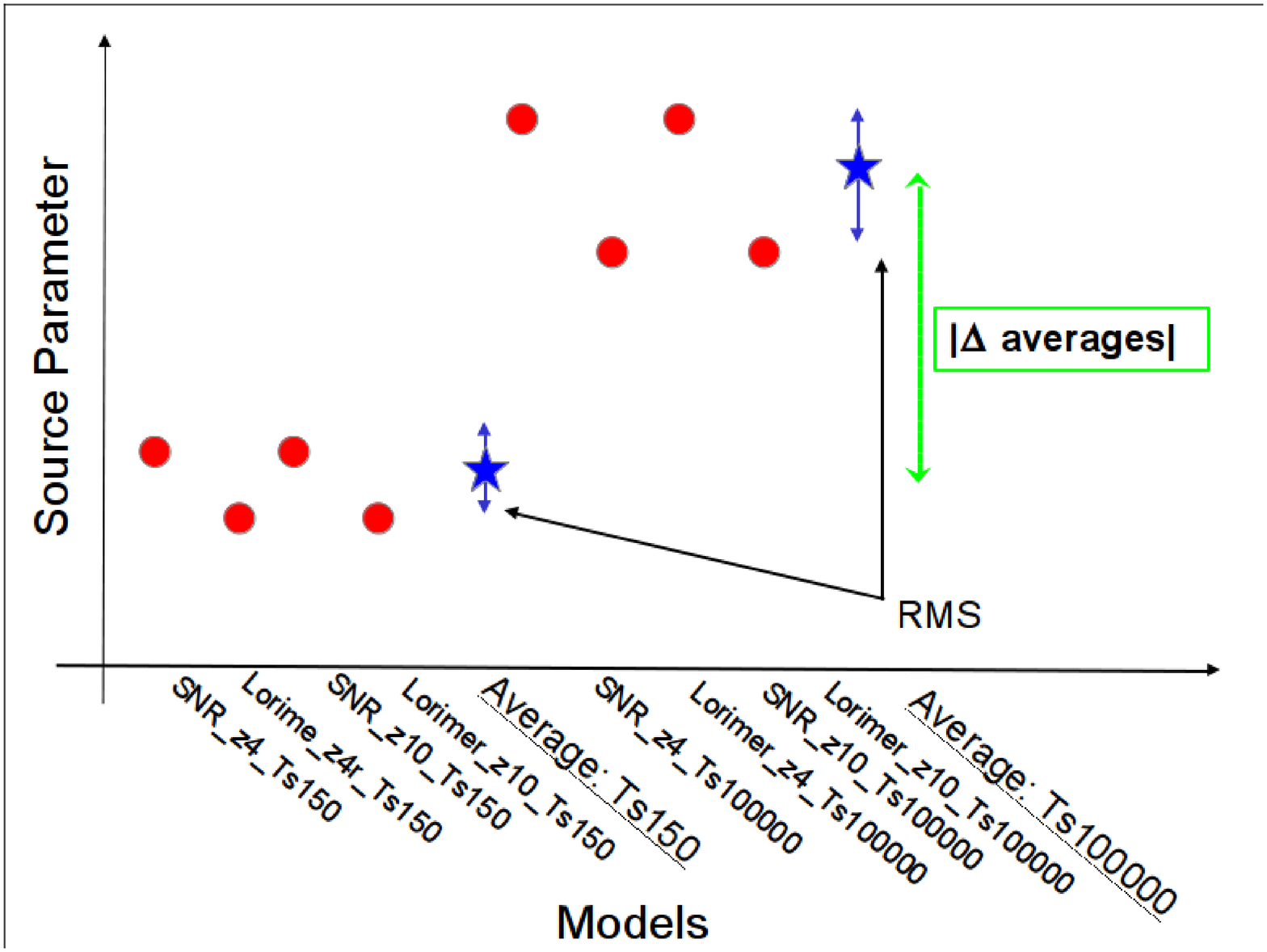}
\caption{Schematic representation of the method for determining the impact of a given GALPROP input parameter (e.g. spin temperature) on a source parameter (e.g. flux) by marginalizing over the other input parameters. We take as the figure of merit the ratio of the difference in averages for $Ts = \{150$\,K$, 10^5$\,K$\}$ divided by the maximum RMS.}
\label{fig:example_IEM} 
\end{figure}

To identify which, if any, of the three IEM input parameters (\hi{} spin temperature, CR source distribution, and CR propagation halo height) has the largest impact on the fitted source parameters, we marginalize over the other parameters and examine the relative ratio of the averaged input parameter values to the values' dispersion. For a fitted source parameter $a$, such as flux and a GALPROP input parameter set $P = \{i, j\}$, e.g. spin temperature $Ts~=~\{150$\,K$, 10^5$\,K$\}$, this becomes:
\begin{equation}
 \frac{|<a_{i}>-<a_{j}>|}{max(\sigma_{a,i},\sigma_{a,j})}
 \label{eq:figMerit}
\end{equation}
where $\sigma_a$ is the rms of the parameter $a$ for a given input parameter value $P$.
Figure~\ref{fig:example_IEM} shows this schematically. A ratio $\geq 1$ implies that changing the selected input parameter has a greater effect on the flux than all combinations of the other input parameters.

\begin{figure*}
        \centering
        \begin{subfigure}[b]{1\columnwidth}
                \centering
                \includegraphics[width=1\columnwidth]{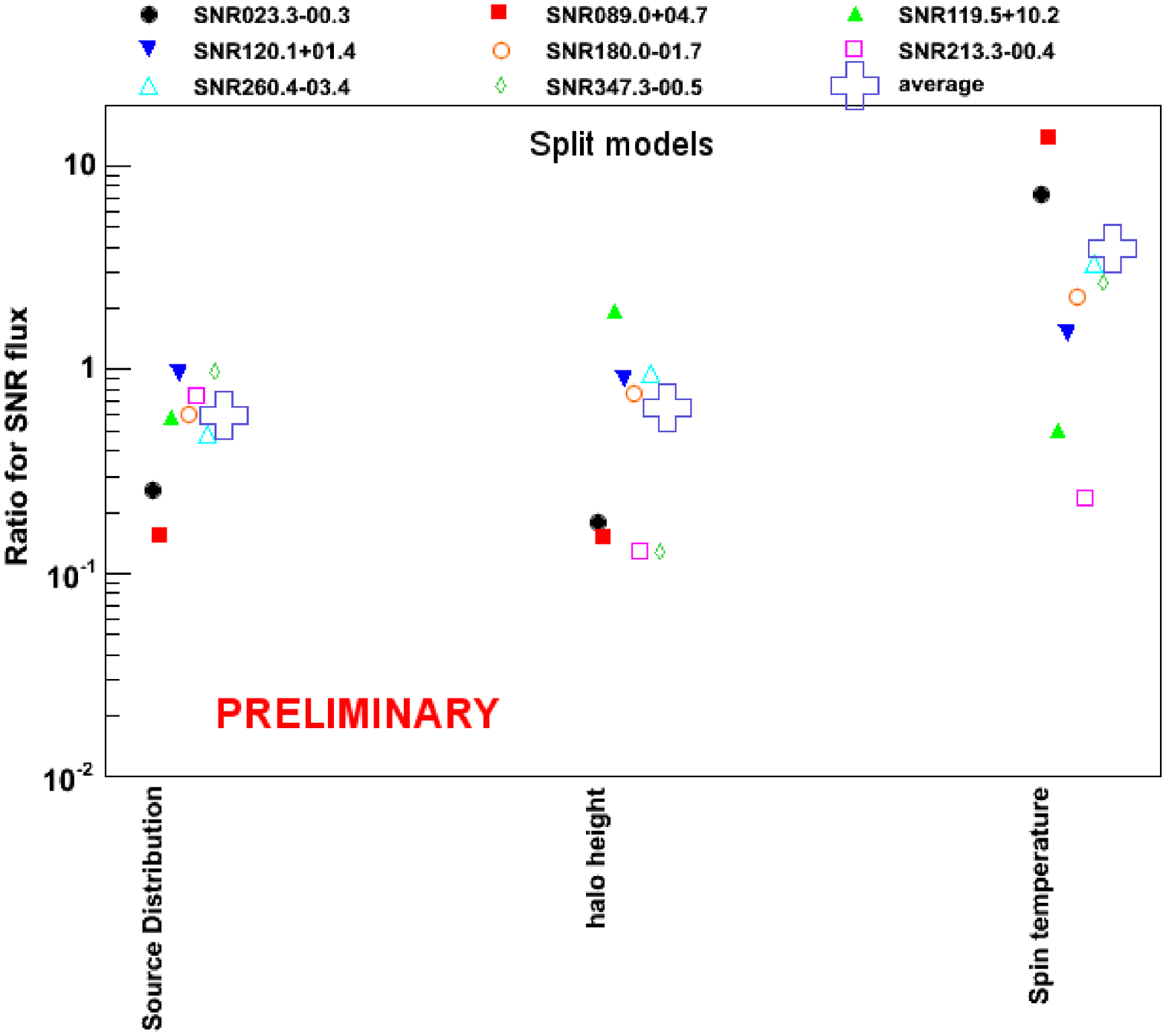}
                \label{fig:IEM_comparison_split}
        \end{subfigure}%
        ~ 
        \begin{subfigure}[b]{1\columnwidth} 
                \centering
                \includegraphics[width=1\columnwidth]{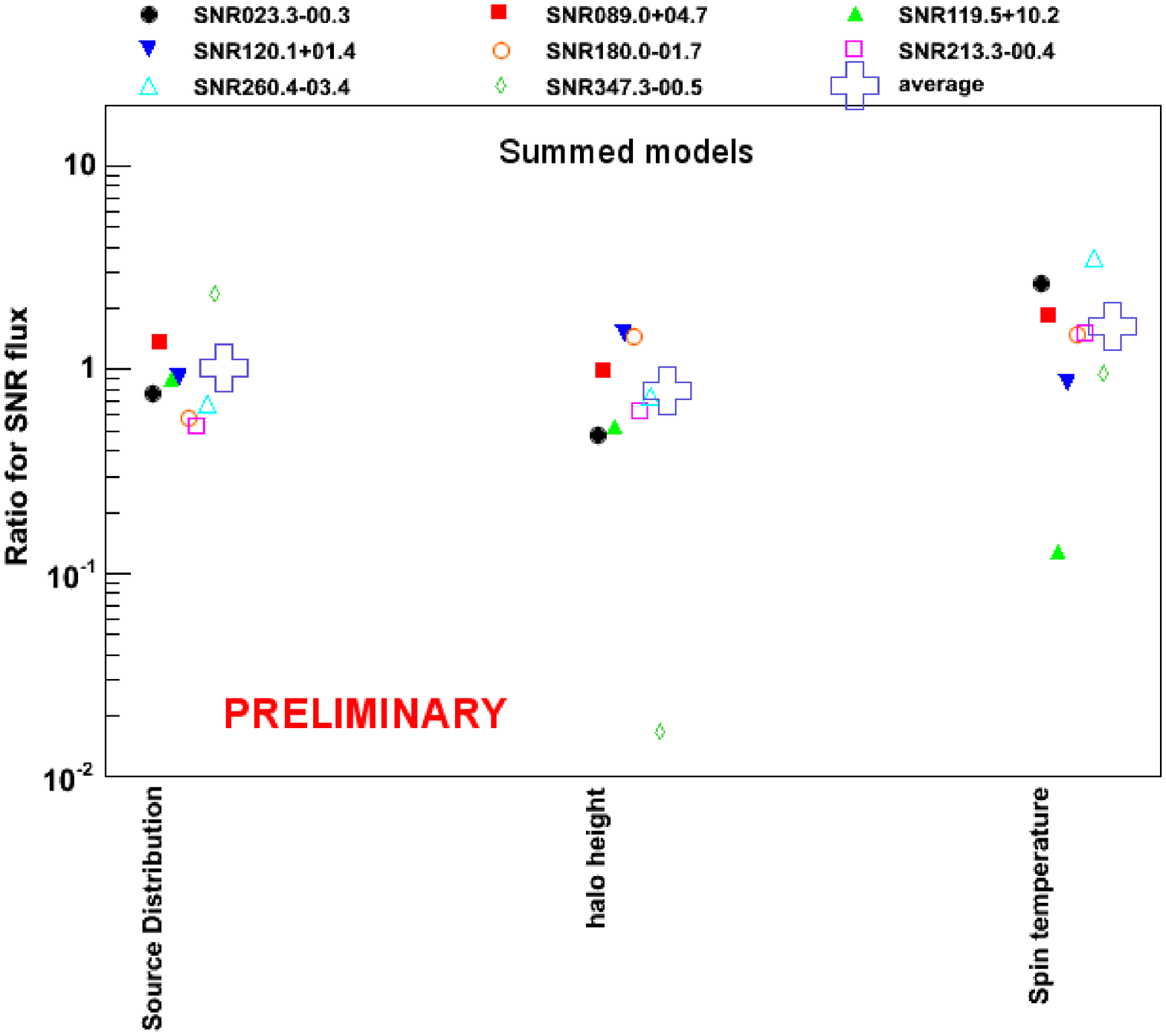}
                \label{fig:IEM_comparison_sum}
        \end{subfigure}
        \caption{The impact on the candidate SNRs' flux of each of the alternative IEM input parameters, marginalized over the other GALPROP input parameters, is shown relative to the figure of merit for the other input parameters (source distribution, halo height, and spin temperature). We calculate the figure of merit (Eq \ref{eq:figMerit}) separately for the alternate IEM components fit separately (left) and summed (right).  The large open cross represents the average figure of merit over all SNR candidates. As no alternative IEM input parameter has a figure of merit significantly larger than $1$, no input parameter dominates the fitted source parameter sufficiently to justify neglecting the others.}
\label{fig:IEM_comparison}
\end{figure*}

In Figure~\ref{fig:IEM_comparison} we plot this ratio for each of the alternative IEM's input parameters for each of the eight SNR candidates, along with the average over the SNR candidates, separately for the split and summed components. While the spin temperature has the largest effect for the split alternative IEMs, the CR source distribution also becomes relevant with the summed alternative models. 
In light of this and as none of the parameters shows a ratio significantly greater than $1$ for all the sources tested, we conclude that none of the input parameters has a sufficiently large impact on the fitted source parameter to justify neglecting the others. 



\section{FUTURE APPLICATIONS}\label{appl} 

In this work we explored the effect of using alternative interstellar emission models on the analysis of LAT sources. As the Galactic interstellar emission contributes substantially to \textit{Fermi} LAT observations in the Galactic plane, the choice of IEM can have a significant impact on the parameters determined for a given source of interest, as demonstrated with eight SNR candidates. To estimate the reported error we currently use only the most conservative extreme variation of the source of interest's output parameters. 
We are finalizing our definition of the systematic error using this method, including through comparison of the present estimate with previous methods' estimates, typically found by varying the standard IEM's normalization by a fraction estimated from neighboring regions. Although our current method represents the uncertainty due to a limited range of IEMs, it plays a critical role in interpreting the data and represents the most complete and systematic attempt at quantifying 
the systematic error due to the choice of IEM to date.

As the majority of SNRs lie in the Galactic plane, coincident with the majority of the Galactic interstellar emission, this method is particularly pertinent to analyses such as that underway for the $1^{st}$ \textit{Fermi} LAT SNR Catalog. Figure~\ref{fig:snr_comparisons} shows that the flux and index can vary greatly for our eight representative SNR candidates, depending on the source and local background's specific characteristics. Given these differences, we plan to use this method to estimate the systematic uncertainty associated with the choice of IEM on the full set of  SNR candidates in the catalog. Such error estimates will allow us to, among other things, more accurately 
determine underlying source characteristics such as the inferred composition (leptonic or hadronic) and particle spectrum. 

Other classes of objects such as pulsar wind nebulae and binary star systems also lie primarily in the plane and are likely to be strongly affected by the choice of IEM. We are thus generalizing this method in order to be able to apply it to the study of Galactic plane sources generally. Another possible extension to this method is extending it to energies $< 1$\,GeV, where the interplay between the Galactic interstellar emission model and background sources must be carefully examined.
By more faithfully accounting for the systematic uncertainty of our model components we will be better equipped to draw less biased conclusions from our data.
\bigskip

\section*{Acknowledgements}
The \textit{Fermi} LAT Collaboration acknowledges generous ongoing support
from a number of agencies and institutes that have supported both the
development and the operation of the LAT as well as scientific data analysis.
These include the National Aeronautics and Space Administration and the
Department of Energy in the United States, the Commissariat \`a l'Energie Atomique
and the Centre National de la Recherche Scientifique / Institut National de Physique
Nucl\'eaire et de Physique des Particules in France, the Agenzia Spaziale Italiana
and the Istituto Nazionale di Fisica Nucleare in Italy, the Ministry of Education,
Culture, Sports, Science and Technology (MEXT), High Energy Accelerator Research
Organization (KEK) and Japan Aerospace Exploration Agency (JAXA) in Japan, and
the K.~A.~Wallenberg Foundation, the Swedish Research Council and the
Swedish National Space Board in Sweden.

Additional support for science analysis during the operations phase is gratefully
acknowledged from the Istituto Nazionale di Astrofisica in Italy and the Centre National d'\'Etudes Spatiales in France.

\bibliographystyle{apsrev.bst}
\bibliography{snr_diffuse}

\end{document}